\documentclass[11pt]{article}
\usepackage{moriond,epsfig}

\def\mco{\multicolumn}

\def\be{\begin{equation}}
\def\ee{\end{equation}}
\def\bea{\begin{eqnarray}}
\def\eea{\end{eqnarray}}

\newcommand{\gev}{\mbox{GeV}}

\newcommand{\eq}[1]{Eq.~(\ref{#1})}
\begin{document}
\vspace*{4cm}
\title{CKM ELEMENTS FROM SQUARK-GLUINO LOOPS}

\author{ ANDREAS CRIVELLIN }

\address{Institut f\"ur Theoretische Teilchenphysik,
Karlsruhe Institute of Technology -- Universit\"at Karlsruhe\\
D-76128 Karlsruhe, Germany}

\maketitle\abstracts{
  I present results for the finite renormalization of the Cabibbo-Kobayashi-Maskawa (CKM)
  matrix induced by gluino--squark diagrams in the MSSM with non-minimal
  sources of flavour violation. Subsequently I derive bounds on the
  flavour--off--diagonal elements of the squark mass matrices by requiring that the radiative corrections to
  the CKM elements do not exceed the experimental values. The constraints on
  the associated dimensionless quantities $\delta^{d\,LR}_{ij}$, $j>i$, are
  stronger than the bounds from flavour-changing neutral current (FCNC)
  processes if gluino and squarks are heavier than 500$\,\gev$. The
  results imply that it is still possible to generate all observed flavour
  violation from the soft supersymmetry-breaking terms without conflicting
  with present-day data on FCNC processes. Therefore we reappraise the idea that only $Y^q_{33}$ 
  is different from zero at tree-level and all other Yukawa couplings are generated radiatively. This model solves the SUSY flavour as well as the SUSY CP problem. The presented results are published in \cite{Crivellin:2008mq}.}

\section{Introduction}

The generic Minimal Supersymmetric Standard Model (MSSM) contains a
plethora of new sources of flavour violation, which reside in the
supersymmetry--breaking sector. Especially the additional flavour 
violation in the squark sector can be dangerously large because 
the squark-quark-gluino vertex, which involves the strong coupling constant, is in general not flavour diagonal. 
This potential failure of the MSSM to describe the small flavour violation observed 
in experiment is known as the "SUSY flavour problem".

We have only partial information from experiment about the quark mass matrices and therefore also about the Yukawa matrices. Not the whole matrices but only their singular values (physical masses) and the misalignment between the rotations of left-handed fields, needed to obtain the mass eigen-basis, (the CKM matrix) are known. This is the reason why it is useful to work in the so called super-CKM basis. We arrive at the super-CKM basis by applying the same rotations which are needed to diagonalize the quark mass matrices to the squark fields:

\begin{equation}
\tilde q^{int} = \left( {\begin{array}{*{20}c}
   {\tilde q_L^{{\mathop{\rm int}} } }  \\
   {\tilde q_R^{{\mathop{\rm int}} } }  \\
\end{array}} \right) \to \tilde{q}^{SCKM}=\left( {\begin{array}{*{20}c}
   {U_L^q } & {0} \\
  {0} & {U_R^q }  \\
\end{array}} \right)\cdot\left( {\begin{array}{*{20}c}
   {\tilde q_L^{{\mathop{\rm int}} } }  \\
   {\tilde q_R^{{\mathop{\rm int}} } }  \\
\end{array}} \right) = \left( {\begin{array}{*{20}c}
   {\tilde q_L^{SCKM} }  \\
   {\tilde q_R^{SCKM} }  \\
\end{array}} \right)
\end{equation}

Here the superscript "int" means interaction eigenstates and the matrices $U_{L,R}^q$ are determined by the requirement that they diagonalize the tree-level quark mass matrices:

\begin{equation}
\renewcommand{\arraystretch}{1.4}
\begin{array}{c}
 U_L^{u\dag } {\bf{m}}_u^{(0)} U_R^{u}  = 
   {\bf{m}}_u^{\left( D \right)},\qquad\qquad
 U_L^{d\dag } {\bf{m}}_d^{(0)} U_R^{d}  = 
{\bf{m}}_d^{\left( D \right)}
 \end{array} \label{defrot}
\end{equation}

In the super-CKM basis the squark mass matrices in the down and in the up sector contain bilinear terms ${\rm M}_{\tilde {q}}^{2}$, ${\rm M}_{\tilde {u}}^{2}$and ${\rm M}_{\tilde {d}}^{2}$ as well as the trilinear terms ${{A}}^{u,d}$ which originate from the soft SUSY breaking and are flavour non-diagonal, in general. All other terms are flavour diagonal in the super-CKM basis and originate from the spontaneous breakdown of ${SU(2)}_L$. Since the squark mass matrices are hermitian they can be diagonalised by a unitary transformations of the squark fields:

\begin{equation}
\tilde{q}^{SCKM}\to\tilde{q}^{mass}=W^{\tilde q}\cdot\tilde{q}^{SCKM},\qquad\qquad M_{\tilde q}^{2\,(D)}=W^{\tilde q \dagger} M_{\tilde q}^2 W^{\tilde q} 
\end{equation}

In the conventions of Ref.~\cite{Gabbiani:1996} the full $6\times 6$ mass matrix is parametrized by

\begin{equation}
\renewcommand{\arraystretch}{1.4}
 M_{\tilde q}^2  = \left( {\begin{array}{*{20}c}
    {\left(M_{1L}^{\tilde d}\right)^2} & {\Delta _{12}^{\tilde{d}\,LL} } & {\Delta
    _{13}^{\tilde{d}\,LL} } & {\Delta _{11}^{\tilde{d}\,LR} } & {\Delta
    _{12}^{\tilde{d}\,LR} } & {\Delta _{13}^{\tilde{d}\,LR} }  \\ 
    {{\Delta _{12}^{\tilde{d}\,LL}}^* } & {\left(M_{2L}^{\tilde d}\right)^2 } & {\Delta
    _{23}^{\tilde{d}\,LL} } & {{\Delta _{12}^{\tilde{d}\,RL}}^* } & {\Delta
    _{22}^{\tilde{d}\,LR} } & {\Delta _{23}^{\tilde{d}\,LR} }  \\ 
    {{\Delta _{13}^{\tilde{d}\,LL}}^* } & {{\Delta _{23}^{\tilde{d}\,LL} }^*} &
    {\left(M_{3L}^{\tilde d}\right)^2 } & {{\Delta _{13}^{\tilde{d}RL}}^* } & {\Delta
    _{23}^{RL*} } & {\Delta _{33}^{\tilde{d}\,LR} }  \\ 
    {{\Delta _{11}^{\tilde{d}\,LR}}^* } & {\Delta _{12}^{\tilde{d}RL} } & {\Delta
    _{13}^{\tilde{d}RL} } & {\left(M_{1R}^{\tilde d}\right)^2 } & {\Delta
    _{12}^{\tilde{d}\,RR} } & {\Delta _{13}^{\tilde{d}\,RR} }  \\ 
    {{\Delta _{12}^{\tilde{d}\,LR}}^* } & {\Delta _{22}^{\tilde{d}\,LR*} } & {\Delta
    _{23}^{\tilde{d}RL} } & {{\Delta _{12}^{\tilde{d}\,RR}}^* } & {\left(M_{2R}^{\tilde
    d}\right)^2 } & {\Delta _{23}^{\tilde{d}\,RR} }  \\ 
    {{\Delta _{13}^{\tilde{d}\,LR}}^* } & {{\Delta _{23}^{\tilde{d}\,LR}}^* } & {{\Delta
    _{33}^{\tilde{d}\,LR}}^* } & {{\Delta _{13}^{\tilde{d}\,RR}}^* } & {{\Delta
    _{23}^{\tilde{d}\,RR}}^* } & {\left(M_{3R}^{\tilde d}\right)^2 }  \\ 
 \end{array}} \right)\label{massmatrix}
\end{equation}

Anticipating the smallness of the off-diagonal elements $\Delta _{ij}^{\tilde q\,XY}$
(with $X,Y=L$ or $R$) it is possible to treat them pertubatively as squark mass terms
~\cite{Gabbiani:1996,Hall:1985dx,Misiak:1997ei,Buras:1997ij}.
It is customary to define the dimensionless quantities

\begin{equation}
\delta^{q \,XY} _{ij}  = 
  \frac{\Delta^{\tilde q\, XY}_{ij}}{\frac{1}{6}\sum\limits_s 
     {\left[M_{\tilde q}^2\right]_{ss}}} .\label{defde}
\end{equation}

Note that the chirality-flipping entries $\delta^{q \,XY} _{ij}$ with $X\neq Y$, even though they are dimensionless, do not stay constant if all SUSY parameters are scaled by a common factor of $a$, but rather decrease like $1/a$.

In the current era of precision flavour physics stringent bounds on the
parameters $\delta^{q \,XY} _{ij}$ have been derived from FCNC processes, by requiring that the
gluino--squark loops do not exceed the measured values of the considered
observables  ~\cite{Gabbiani:1996,Hagelin:1992,Ciuchini:1998ix,Borzumati:1999,Becirevic:2001,Silvestrini:2007,Ciuchini:2007cw}

We will show in the next section that even more stringent bounds on these quantities can be obtained if we apply a fine-tuning argument which assumes the absence of large accidental cancellations between different contributions to the CKM matrix.

\section{Renormalization of the CKM matrix}

The simplest diagram (and at least for our discussion the most important one) in which this new flavour and chirality violations induced by the squark mass matrices enters is a potentially flavour-changing self-energy with a squark and a gluino as virtual particles. Since the SUSY particles are much heavier than the five lightest quarks,
it is possible in the calculation of these diagrams to expand in the external momentum, unless one external
quark is the top. In the following we consider the self-energies with
only light external quarks. The case with a top quark as an external quarks is discussed in ~\cite{Crivellin:2008mq}. Direct computation of the diagram gives:

\begin{eqnarray}
\Sigma^{q\,LR}_{fi} (p^2=0) =
\frac{{2m_{\tilde g} }}{{3\pi }}\alpha _s (M_{\rm SUSY})
      \sum_{s = 1}^6 
      W_{f + 3,s}^{\tilde q} W_{is}^{\tilde{q}*}     
  B_0 \left( {m_{\tilde g} ,m_{\tilde q_s } } \right) \\
  \Sigma^{q\,RL}_{fi} (p^2=0) =
\frac{{2m_{\tilde g} }}{{3\pi }}\alpha _s (M_{\rm SUSY})
      \sum_{s = 1}^6 
      W_{f,s}^{\tilde q} W_{i+3,s}^{\tilde{q}*}     
  B_0 \left( {m_{\tilde g} ,m_{\tilde q_s } } \right)
\label{selbstenergie}
\end{eqnarray}

For our definition of the loop-function $B_0$ see appendix of ~\cite{Crivellin:2008mq}. This self-energy has several important properties:

\begin{itemize}
  \item It is finite and independent of the renormalization scale.
	\item It is always chirality-flipping. 
	\item It does not decouple but rather converges to a constant if all SUSY parameter go to infinity.
	\item It satisfies $\Sigma^{q\,LR}_{fi}=\Sigma^{q\,RL\,*}_{if}$.
	\item It is chirally enhanced by an approximate factor of $\frac{\left|A^q_{fi}\right|}{M_{SUSY} \left|Y^q_{fi}\right|}$ or $\frac{v \tan\beta}{M_{SUSY}}$ compared to the tree-level quark coupling. These factors may compensate for the loop suppression factor of $1/(16 \pi^2)$.
\end{itemize}

In the case when this self-energy is flavour conserving, it renormalizes the corresponding quark mass in a rather trivial way:

\begin{equation}
m_{q_i}^{(0)}\to m_{q_i}=m_{q_i}^{(0)}+\Sigma^{q\,LR}_{ii}
\end{equation}

Since the self-energy is finite the introduction of a counter-term is optional. In minimal renormalization schemes the counter-term is absent and in the one-shell scheme it just equals $-\Sigma^{q\,LR}_{ii}$. Since we will later consider the possibility that the light quark masses are generated exclusively via these loops, meaning $m_{q_i}=\Sigma^{q\,LR}_{ii}$, it is most natural and intuitive to choose a minimal renormalization scheme like $\overline{\rm MS}$.

\begin{figure}
\includegraphics[width=1\textwidth]{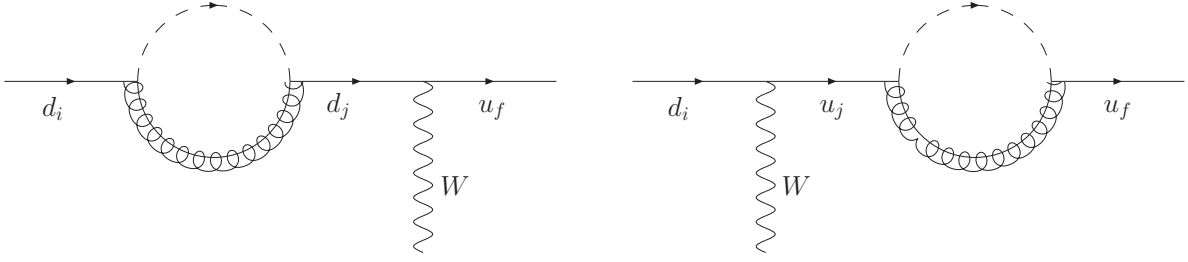}
\caption{One-loop corrections to the CKM matrix from the down and 
  up sectors contributing to $\Delta U_L^d$ and $\Delta U_L^u$ in \eq{physv}, respectively. \label{fig:W}}
\end{figure}

The renormalization of the CKM matrix is a bit more involved. There are two possible contributions, the self-energy
diagrams and the proper vertex correction. The vertex diagrams
involving a $W$ coupling to squarks are not chirally enhanced and moreover suffer
from gauge cancellations with non-enhanced pieces from the
self-energies. Therefore we only need to consider self-energies, just as
in the case of the electroweak renormalization of $V$ in the SM
~\cite{Denner:1990}. The two diagrams shown in Fig 1 contribute at the one loop level. According to \cite{Logan:2000iv} they can be treated in the same way as one-particle-irreducible vertex corrections. Computing theses diagrams we receive the following corrections to the CKM matrix:

\begin{equation}
        V^{(0)} \to V  =   \left(1+\Delta
          U_L^{u\dag}\right)V^{(0)}\left(1+\Delta U_L^d\right) 
 \label{physv}
\end{equation}

with

\begin{equation}
\renewcommand{\arraystretch}{1.4}
\Delta U_L^q \,=\, 
\left( {\begin{array}{*{6}c}
0 & 
{\frac{1}{{m_{q_2 }}} {\Sigma _{12}^{q\,LR} }  } & 
{\frac{1}{{m_{q_3 }}} {\Sigma _{13}^{q\,LR} } }  \\
           {\frac{{ - 1}}{{m_{q_2 }}} {\Sigma _{21}^{q\,RL} }  } & 
0 & 
{\frac{1}{{m_{q_3 }}} {\Sigma _{23}^{q\,LR} } }  \\
           {\frac{{ - 1}}{{m_{q_3 }}} {\Sigma _{31}^{q\,RL} } } & 
{\frac{{ - 1}}{{m_{q_3 }}}{\Sigma _{32}^{q\,RL} }  } & 0
        \end{array}} \right)
\label{DeltaU} 
\end{equation}

In \eq{DeltaU} we have discarded small quark-mass ratios. Just as in the case of the mass-renormalization we choose a minimal renormalization scheme which complies with the use of the super-CKM basis (see \cite{Crivellin:2008mq} for details). It is easily seen from \eq{DeltaU} that the corrections are antihermitian which is in agreement with the demanded unitarity of the CKM matrix at one-loop. Our corrections are independent of the renormalization scale $\mu$. The choice $\mu=M_{SUSY}$ avoids large logarithms in $\Sigma_{ij}^{q\,LR}$, so that we have to evaluate it at this scale. This means we must also evaluate the quark masses appearing in \eq{DeltaU} at the scale $M_{SUSY}$.

We can now receive constraints on the off-diagonal elements of the squark
mass matrices from \eq{physv} by applying a fine-tuning argument.
Large accidental cancellations between the SM and supersymmetric
contributions are, as already mentioned in the introduction, unlikely
and from the theoretical point of view undesirable. Requiring the
absence of such cancellations is a commonly used fine-tuning argument,
which is also employed in standard FCNC analyses of the $\delta^{q \,XY}
_{ij}$'s ~\cite{Gabbiani:1996,Hagelin:1992,Ciuchini:1998ix,Borzumati:1999,Becirevic:2001,Silvestrini:2007,Ciuchini:2007cw}.  Analogously, we
assume that the corrections due to flavour-changing SQCD self-energies
do not exceed the experimentally measured values for the CKM matrix
elements quoted in the Particle Data Table (PDT) ~\cite{Amsler:2008zzb}. To this
end we set the tree-level CKM matrix $V^{(0)}$ equal to the unit matrix and generate the measured values
radiatively. For $m_{\tilde q}=m_{\tilde g}=1000 \gev$ we receive the constraints quoted in table 1.

\begin{table}[t]
\caption{Comparisons of our constraints on $\delta^{q \,XY} _{ij}$ with the constraints obtained from FCNC processes and vacuum stability bounds.}
\vspace{0.6cm}
\begin{center}
\renewcommand{\arraystretch}{1.3}
\begin{tabular}{|c|l|l l|l|}
\hline

quantity   & our bound   & \mco{2}{|c|}{bound from FCNC's} & bound from VS ~\cite{Casas:1995}\\ \hline

$|\delta^{d\,LR}_{12} |$ & $\leq 0.0011$   & $\leq 0.006$ & 
$K$ mixing ~\cite{Ciuchini:1998ix}   & $\leq 1.5\, \times \,10^{-4}$  \\
$|\delta^{d\,LR}_{13} |$ & $\leq 0.0010$   & $\leq 0.15$ & 
$B_d$ mixing ~\cite{Becirevic:2001}   & $\leq 0.05$  \\
$|\delta^{d\,LR}_{23} |$ & $\leq 0.010$        & $\leq 0.06$ & 
 $B\rightarrow X_s\gamma; X_s l^+l^-$ ~\cite{Silvestrini:2007}   & 
$\leq 0.05$  \\
$|\delta^{d\,LL}_{13} |$ & $\leq 0.032$ & $\leq 0.5$ & 
$B_d$ mixing ~\cite{Becirevic:2001}    & $-$  \\
$|\delta^{u\,LR}_{12} |$ & $\leq 0.011$  & $\leq 0.016$  & 
$D$ mixing ~\cite{Ciuchini:2007cw}  & $\leq 1.2\, \times \,10^{-3}$  \\
$|\delta^{u\,LR}_{13} |$ & $\leq 0.062$       & \mco{2}{|c|}{---}
& $\leq 0.22$  \\
$|\delta^{u\,LR}_{23} |$ & $\leq 0.59$        & \mco{2}{|c|}{---}
& $\leq 0.22$  \\ \hline                    
\end{tabular}
\end{center}
\end{table}

Note that our constraints are all much stronger than the FCNC bounds. The FCNC bounds in addition decouple. This means they vanish like $1/a^2$ if all SUSY masses are scaled with $a$. The vacuum stability bounds are stronger than our ones for the $\delta^{q\,LR}_{12}$ elements and they are non-decoupling like our bounds. However, the analysis of ~\cite{Casas:1995} only takes tree-level Yukawa coupling into account and the small Yukawa couplings are modified by the very same loop effects which enter $\Delta U_{L,R}^q$ in \eq{DeltaU}.

\section{The Model}

The smallness of the Yukawa couplings of the first two generations suggests that these
couplings are generated through radiative corrections \cite{Weinberg:1972ws}.
In the context of supersymmetric theories these loop-induced couplings arise
from diagrams involving squarks and gluinos. Although the B factories have confirmed the CKM mechanism of flavour violation with very high precision, leaving little room for new sources of FCNCs, the possibility of radiative generation of quark masses and of the CKM matrix still remains valid as proven in section 2, even for SUSY masses well below 1 TeV, if the sources of flavour violation are the trilinear terms \cite{Crivellin:2008mq}.
Of course, the heaviness of the top quark requires a special treatment of $Y^t$ and the successful bottom-tau Yukawa unification suggests to keep tree-level Yukawa couplings for the third generation.
At large $\tan \beta$, this idea gets even more support from the successful unification of the top and bottom Yukawa coupling, as suggested by some GUT models. Radiative Yukawa interactions from SUSY-breaking terms have been considered earlier in Refs \cite{Buchmuller:1982ye,Ferrandis:2004,Borzumati:1997bd}.

In the modern language of Refs.~\cite{D'Ambrosio:2002ex,cg} the global $[U(3)]^3$ flavour
symmetry of the gauge sector (here we do not consider neutrinos) is broken to $[U(2)]^3 \times [U(1)]$ by the Yukawa couplings of the third
generation. Here the three $U(2)$ factors correspond to rotations of the
left-handed doublets and the right-handed singlets of
the first two generation quarks in flavour space, respectively. 

This means we have

\begin{equation}
Y^{q}  = \left( {\begin{array}{*{20}c}
   0 & 0 & 0  \\
   0 & 0 & 0  \\
   0 & 0 & y^q  \\
\end{array}} \right),\;\;\;V^{(0)}  = \left( {\begin{array}{*{20}c}
   1 & 0 & 0  \\
   0 & 1 & 0  \\
   0 & 0 & 1  \\
\end{array}} \right)
\end{equation}

in the tree-level Lagrangian. 
We next assume that the soft breaking terms ${\Delta _{ij}^{\tilde{q}\,LL} }$ and ${\Delta _{ij}^{\tilde{q}\,RR} }$
possess the same flavour symmetry as the Yukawa sector, which implies that ${\bf{M}}_{\tilde q}$ , ${\bf{M}}_{\tilde d}$ and ${\bf{M}}_{\tilde u}$ are diagonal matrices with the first two entries being equal. For transitions involving the third generation the situation is different because flavour violation can occur not only because of a misalignment between $A^u$ and $A^d$ but also due to a misalignment with the Yukawa matrix. So the elements $A^{u,d}_{j3}$ do not only generate the CKM matrix at one-loop, they also act as a source of non-minimal flavour violation and thus can be constrained by FCNC processes.

This model has several advantages compared with the generic MSSM:
\begin{itemize}
	\item Flavour universality holds for the first two generations. Thus our Model is minimally flavour violating according to the definition of \cite{D'Ambrosio:2002ex} with respect to the first two generations since the quark and the squark mass matrices are diagonal in the same basis. This provides a explanation of the precise agreement between theory and experiment in K and D physics.
	\item The SUSY flavour problem is reduced to the quantities $\delta^{q\,RL}_{13,23}$. However, these flavour-changing elements are less constrained from FCNCs and might even explain a possible new CP phase indicated by recent data on $B_s$ mixing.
	\item The flavour symmetry of the Yukawa sector protects the quarks of the first two generations from a tree-level mass term.
	\item The model is economical: Flavour violation and SUSY breaking have the same origin. Small quark masses and small off-diagonal CKM elements
are explained by a loop suppression.
\item The SUSY CP problem is substantially alleviated by an automatic phase alignment \cite{Borzumati:1997bd}. In addition, the phase of $\mu$ does not enter the EDMs at the one-loop level, because
the Yukawa couplings of the first two generations are zero. 
\end{itemize}

\section{Conclusions}
We have computed the renormalization of the CKM matrix by chirally-enhanced
flavour-changing SQCD effects in the MSSM with generic flavour structure \cite{Crivellin:2008mq}. We have derived upper bounds on the
flavour-changing off-diagonal elements $\Delta _{ij}^{\tilde{q}\,XY}$ of the
squark mass matrices by requiring that the supersymmetric corrections do not
exceed the measured values of the CKM elements.  For $M_{\rm SUSY}\geq
500\,\gev$ our constraints on \emph{all}\ elements $\Delta
_{ij}^{\tilde{d}\,LR}$, $i<j$, are stronger than the constraints from FCNC
processes. As an important consequence, we conclude that it is possible to generate the
observed CKM elements completely through finite supersymmetric loop diagrams
\cite{Buchmuller:1982ye,Ferrandis:2004} without violating present-day data on
FCNC processes. In this scenario the Yukawa sector possesses a higher flavour
symmetry than the trilinear SUSY breaking terms. Additional applications to 
charged Higgs and chargino couplings are considered in \cite{Crivellin:2008mq}.

\section*{Acknowledgments}

This work is supported by BMBF grant 05 HT6VKB and by the EU Contract
No.~MRTN-CT-2006-035482, \lq\lq FLAVIAnet''. I am grateful to the organizers for inviting me to this conference. I like to thank Lars Hofer for reading the manuscript and many useful discussions. I am grateful to Ulrich Nierste for the collaboration on the presented work ~\cite{Crivellin:2008mq}.

\end{document}